\documentclass{moriond}

\usepackage{amsfonts}
\usepackage{amsmath}
\usepackage{amssymb}
\usepackage{mathtools}
\usepackage{subcaption}
\usepackage{tikz}
\usepackage{tikz-feynman}
\usetikzlibrary{decorations.markings}
\usepackage{graphicx}  % got figures? uncomment this

\newcommand{\colordiagram}[1]{
\begin{subfigure}[b]{0.31\textwidth}
\begin{tikzpicture}
\begin{feynman}
#1
\end{feynman}
\end{tikzpicture}
\end{subfigure}
}
\def\fermions{
    \vertex [] (psi1) {\footnotesize $\overline{\Psi}_L$};
    \vertex [right=0.8in of psi1] (c0);
    \vertex [right=1.6in of psi1] (psi2) {\footnotesize $\Psi_R$};
    \vertex [above left =0.05in and 0.05in of c0.east] (L0);
    \vertex [above right =0.05in and 0.05in of c0.west] (R0);
    \vertex [above  =0.05in of psi1.east] (psi1V);
    \draw[red]  (L0)--(psi1V);
    \vertex [above  =0.05in of psi2.west] (psi2V);
    \draw[blue,dash pattern={on 3pt off 1pt}]  (psi2V)--(R0);
    \diagram* {
    (psi1) -- [fermion] (c0)  -- [anti fermion] (psi2)
    };
}
\def\sep{0.35in}

\newcommand{\HiggsII}[2]{
    \vertex [above=\sep of c0] (c1);
    \vertex [above=\sep of c1] (c2) {\footnotesize $\langle H_a\rangle$};
    \diagram* {(c0)--[scalar] (c1)--[scalar] (c2)};
    \vertex [ left  =0.05in of c2.south] (#1);
    \vertex [ right  =0.05in of c2.south] (#2);
}

\newcommand{\phiLI}[1]{
    \vertex [left=0 in of #1] (A);
    \vertex [below left =0.02in and 0.05in of A.west] (L1);
    \vertex [above left =0.02in and 0.05in of A.west] (L2);
    \vertex [left=0.3in of A] (E) {\footnotesize $\langle\Phi_L\rangle$};
    \diagram* {(A) --[scalar] (E)};
    \vertex [left=0.27in of L1] (B);
    \draw[red] (B)--(L1);
    \vertex [left=0.27in of L2] (D);
    \draw[red] (L2)-- (D);
}

\newcommand{\phiRI}[1]{
    \vertex [right=0 in of #1] (A);
    \vertex [below right =0.02in and 0.05in of A.east] (R1);
    \vertex [above right =0.02in and 0.05in of A.east] (R2);
    \vertex [right=0.3in of A] (E) {\footnotesize $\langle\Phi_R\rangle$};
    \diagram* {(A) --[scalar] (E)};
    \vertex [right=0.27in of R1] (B);
    \draw[blue,dash pattern={on 3pt off 1pt}] (B)--(R1);
    \vertex [right=0.27in of R2] (D);
    \draw[blue,dash pattern={on 3pt off 1pt}] (D)-- (R2);
}

\newcommand{\LIII}[2]{
\draw[red,postaction={decoration={markings,mark=at position 0.6 with {\arrow{<<<}}},decorate}]  (#1)-- (#2);
}
\newcommand{\LII}[2]{
\draw[red,postaction={decoration={markings,mark=at position 0.5 with {\arrow{<<}}},decorate}]  (#1)-- (#2);
}

\newcommand{\RIII}[2]{
\draw[blue,dash pattern={on 3pt off 1pt},postaction={decoration={markings,mark=at position 0.6 with {\arrow{>>>}}},decorate}]  (#1)-- (#2);
}
\newcommand{\RII}[2]{
\draw[blue,dash pattern={on 3pt off 1pt},postaction={decoration={markings,mark=at position 0.5 with {\arrow{>>}}},decorate}]  (#1)-- (#2);
}

\def\g{\mathfrak{g}}

\def\u{\mathfrak{u}(1)}

\def\so{\mathfrak{so}}
\def\su{\mathfrak{su}}
\def\sp{\mathfrak{sp}}

\newcommand{\Photo}{}

\title{Gauge flavour unification: \\
from  the flavour puzzle to stable protons}
\author{Joe Davighi}
\address{Physik-Institut, Universit\"at Z\"urich, CH-8057 Z\"urich, Switzerland}
\begin{document}

\maketitle

\begin{abstract}
The idea of unification attempts to explain the structure of the Standard Model (SM) in terms of fewer fundamental forces and/or matter fields. However, traditional grand unified theories based on $SU(5)$ and $\mathrm{Spin}(10)$ shed no light on the existence of three generations of fermions, nor the distinctive pattern of their Yukawa couplings to the Higgs. We discuss two routes for unifying the SM gauge symmetry with its flavour symmetries: firstly, unifying flavour with electroweak symmetries via the group $SU(4) \times Sp(6)_L \times Sp(6)_R$; secondly, unifying flavour and colour via $SU(12) \times SU(2)_L \times SU(2)_R$. In either case, all three generations of SM fermions are unified into just two fundamental fields. In the larger part of this proceeding, we describe how the former model of `electroweak flavour unification' offers a new explanation of hierarchical fermion masses and CKM angles. As a postscript, we show that gauge flavour unification can have unexpected spin-offs not obviously related to flavour. In particular, the $SU(12) \times SU(2)_L \times SU(2)_R$ symmetry, when broken, can leave behind remnant discrete gauge symmetries that exactly stabilize protons to all orders.
\end{abstract}

\section{A unified theory of flavour}

The Standard Model (SM), while familiar, is a complicated quantum field theory. The gauge algebra $G_{\text{SM}}:=\su(3)\oplus \su(2)_L\oplus \u_Y$ entails three independent gauge couplings, corresponding to both non-abelian and abelian forces. There are five (or six, if including right-handed neutrinos) fermion fields transforming in a curious set of irreducible representations (henceforth {\em irreps}), a structure that is repeated across three generations. The generations are themselves a source of yet more unexplained structure, though the distinctive pattern of their Yukawa couplings to the Higgs field. The dynamics of the SM are just as rich; the Higgs mechanism gives rise to weak, short-range forces as well as a long-range electromagnetic force, while QCD becomes strongly coupled and confines. Unification of forces and/or matter attempts to explain all (or part of) this structure as a consequence of something simpler at high energies.

%\begin{figure}
%\includegraphics{foo}     % includes figure foo.eps
%\caption{Description of the figure.}
%\end{figure}

Famously, there are just two grand unified theories that don't require extra fermions (beyond a right-handed neutrino), based on the gauge algebras $\su(5)$~\cite{Georgi:1974sy} and $\so(10)$~\cite{Fritzsch:1974nn,Georgi:1974my}. The SM fermions, including a right-handed neutrino, transform in the representations $[{\bf 5 \oplus \overline{10} \oplus 1}]^{\oplus 3}$ and ${\bf 16}^{\oplus 3}$ respectively. But these models shed no light on flavour. Intriguingly, there is no simple gauge algebra that further packages together the three generations into a gauge-flavour unified model (at least not without requiring extra fermions). It seems that, if one wishes to unify the three generations of matter, one must forgo the complete unification of forces.

Here we explore possible routes for unifying the gauge and flavour symmetries of the SM~\cite{Davighi:2022fer,Davighi:2022qgb}. As proven by exhaustion in ref.~\cite{Allanach:2021bfe}, there are three semi-simple gauge algebras that unify all the fermions of the SM+$3\nu_R$ into just two representations. %, the smallest possible number.\footnote{We ignore trivial extensions of $\so(10)$ by anomaly-free horizontal family symmetries.} 
All three are flavour enrichments of the quark-lepton-unifying $\mathfrak{ps}:=\su(4) \oplus \su(2)_L \oplus \su(2)_R$ of Pati--Salam~\cite{Pati:1974yy}, for which the fermions transform as $\psi_{L} \sim ({\bf 4, 2, 1})^{\oplus 3}$ and $\psi_R \sim({\bf 4, 1, 2})^{\oplus 3}$. The first option is to unify electroweak symmetry with a chiral $\so(3)_L \oplus \so(3)_R$ flavour symmetry. This can be done in two ways, either preserving left-right symmetry with $\g_{\mathrm{EWF}}:=\su(4)\oplus \sp(6)_L \oplus \sp(6)_R$ (see also~\cite{Kuo:1984md}), or not, with $\g_{\mathrm{EWF'}}:=\su(4)\oplus \sp(6)_L \oplus \so(6)_R$. Here, $\sp(6)$ denotes the Lie algebra of $Sp(6)$, which is the Lie subgroup of $SU(6)$ matrices $U$ that further satisfy $U^T \Omega U = \Omega$, where $\Omega = \begin{psmallmatrix} 0 & I_3 \\ -I_3 & 0 \end{psmallmatrix}$. In both these cases, the fermions transform in the two representations $\psi_L \sim({\bf 4, 6, 1})$ and $\psi_R \sim ({\bf 4, 1, 6})$. The second option is to unify the $\su(4)$ `colour' symmetry with a complex $\su(3)$ family symmetry, to arrive at $\g_{\text{CF}}:=\su(12)\oplus \su(2)\oplus \su(2)$, with fermions transforming as  a left-handed $({\bf 12, 2, 1})$ plus a right-handed $({\bf 12, 1, 2})$.

Each of the three gauge algebras $\g_{\mathrm{CF}}$, $\g_{\mathrm{EWF}}$ or $\g_{\mathrm{EWF'}}$ intricately intertwines the flavour symmetry with the SM gauge symmetries, unifying all three families together in the same multiplet, thereby explaining why there are three families in terms of an underlying gauge symmetry. In this proceeding, we further show that electroweak flavour unification~\cite{Davighi:2022fer}, based on $\g_{\mathrm{EWF}}$, can provide an elegant model for the hierarchical structure of fermion masses and CKM mixing angles, with no ingredients needed beyond a set of scalars that break the symmetry down to the SM. In sect. \ref{sec:colour} we briefly discuss the other route, namely colour flavour unification, remarking that this symmetry contains discrete gauge subgroups $\Gamma \cong \mathbb{Z}_{3n_f} = \mathbb{Z}_{9}$ that conserve baryon number mod $n_g$, serving to exactly stabilize the proton to all orders~\cite{Davighi:2022qgb} when the number of generations $n_g \geq 2$.

\section{Electroweak flavour unification}

\subsection{Embedding the SM}

We here describe the basics of the electroweak flavour unification model, based on the anomaly-free UV gauge group
\begin{equation}
G_{\mathrm{EWF}}=SU(4)\times Sp(6)_L \times Sp(6)_R\, ,
\end{equation}
the Lie algebra of which is $\g_{\mathrm{EWF}}$.
The SM chiral fermions are embedded in two fields. All 24 left-handed Weyl fermions are embedded in the left-handed field
\begin{equation}
\Psi_L \sim ({\bf 4, 6, 1}) \sim \begin{pmatrix}
u_1^r & u_2^r & u_3^r & d_1^r & d_2^r & d_3^r\\
u_1^g & u_2^g & u_3^g & d_1^g & d_2^g & d_3^g\\
u_1^b & u_2^b & u_3^b & d_1^b & d_2^b & d_3^b\\
\nu_1 & \nu_2 & \nu_3 & e_1 & e_2 & e_3\\
\end{pmatrix}\, ,
\end{equation}
where the subscript indices indicate family number.
All 24 right-handed fields are similarly embedded in a field $\Psi_R \sim ({\bf 4, 1, 6})$.

Recall that family-blind models with $SU(4)$ quark lepton unification, such as those based on the Pati--Salam group, typically require two Higgs fields, in the singlet and adjoint of $SU(4)$, in order to split quark and lepton masses.  The same is true here except now the electroweak indices of each Higgs field must extend to $Sp(6)$ indices, meaning that the Higgs necessarily picks up flavour quantum numbers. The minimal embedding of Higgs fields is in representations $H_1 \sim ({\bf 1, 6, 6})$ and  $H_{15} \sim ({\bf 15, 6, 6})$.
We can write down renormalisable Yukawa couplings with these fields, 
\begin{equation} \label{eq:yuk}
\mathcal{L} = \sum_{a\in\{1,15\}}
y_a \mathrm{Tr} [\overline{\Psi}_L \Omega H_a \Omega^T \Psi_R] + 
%y_{15} \mathrm{Tr} [\overline{\Psi}_L \Omega H_{15} \Omega \Psi_R] + 
\overline{y}_a \mathrm{Tr} [\overline{\Psi}_L \Omega H_a^\ast \Omega^T \Psi_R] \, ,
%\overline{y}_{15} \mathrm{Tr} [\overline{\Psi}_L \Omega H_{15}^\ast \Omega \Psi_R]\, .
\end{equation}
where recall $\Omega$ denotes the constant matrix $\Omega = \begin{psmallmatrix} 0 & I_3 \\ -I_3 & 0 \end{psmallmatrix}$ throughout.

The symmetry $G_{\mathrm{EWF}}$ must ultimately be broken down to $G_{\mathrm{SM}}=SU(3)\times SU(2)_L\times U(1)_Y$. We do so using an (almost) minimal set of scalars $S_L$, $S_R$, $\Phi_L$ and $\Phi_R$, whose quantum numbers are recorded in Table~\ref{tab:fields} which summarizes the entire field content of the model. The scalars $S_L$ and $\Phi_L$ are real, while $S_R$ and $\Phi_R$ are complex.

\begin{table}[h]
\begin{center}
\begin{tabular}{|c|c|c|}
\hline
Type & Field & $G_{\mathrm{EWF}}$ irrep \\
\hline
\hline
%\hline
Standard Model fermions & $\Psi_L$ & $(\mathbf{4,6,1})$  \\
& $\Psi_R$ & $(\mathbf{4,1,6})$ \\
\hline
%\hline
Higgs fields & $H_1$ & $(\mathbf{1,6,6})$ \\
& $H_{15}$ & $(\mathbf{15,6,6})$ \\
\hline
%\hline
Symmetry breaking scalars & $S_L$ ($\mathbb{R}$) & $(\mathbf{1,14,1})$ \\
& $S_R$ ($\mathbb{C}$) & $(\mathbf{\overline{4},1,6})$ \\
& $ \Phi_L$ ($\mathbb{R}$) & $(\mathbf{1,14,1})$ \\
& $\Phi_{R}$ ($\mathbb{C}$) & $(\mathbf{1,1,14})$ \\
\hline
\end{tabular}
\end{center}
\caption{{\small Field content of the electroweak flavour unification model.
For the symmetry breaking scalars, we also indicate whether they are real ($\mathbb{R}$) or complex ($\mathbb{C}$) scalars.
}} \label{tab:fields}
\end{table}

\subsection{Symmetry breaking pattern} \label{sec:SSB}

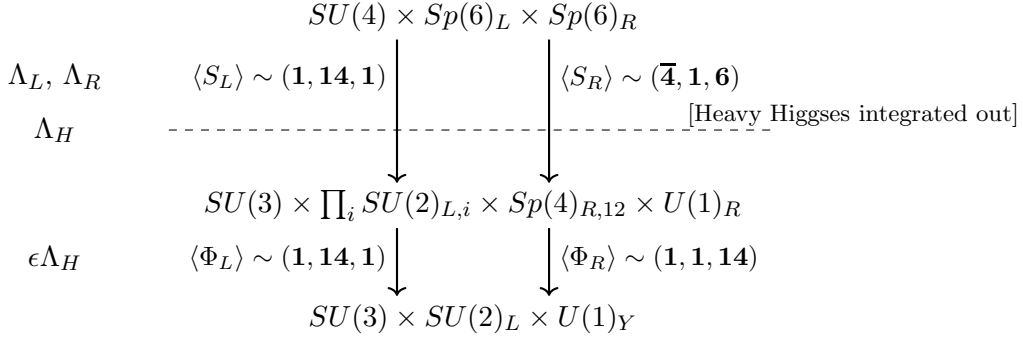
\begin{figure}[t]
\begin{center}
\begin{tikzpicture}
%%Gauge Symmetries
\node at (0,4){$SU(4)\times Sp(6)_L \times Sp(6)_R$};
\node at (0,1.5){$SU(3)\times \prod_i SU(2)_{L,i} \times Sp(4)_{R,12}\times U(1)_{R}$};
\node at (0,0){$SU(3)\times SU(2)_L \times U(1)_Y$};
%%Scales
\node at (-5.5,3.2){$\Lambda_L$, $\Lambda_R$};
\node at (-5.5,2.5){$\Lambda_H$};
\node at (5,2.7){\footnotesize{[Heavy Higgses integrated out]}};
\node at (-5.5,0.8){$\epsilon \Lambda_H$};
%%EFT matching line
\draw[dashed] (-4,2.5)--(4,2.5);
%%UV to int arrows
\draw[->,thick] (-1,3.7)--(-1,1.8);
\node[anchor=east] at (-1,3.2){{\small $\langle S_L \rangle \sim ({\bf 1}, {\bf 14}, {\bf 1})$}};
\draw[->,thick] (1,3.7)--(1,1.8);
\node[anchor=west] at (1,3.2){{\small $\langle S_R \rangle \sim (\overline{\bf 4}, {\bf 1}, {\bf 6})$}};
%%int to IR arrows
\draw[->,thick] (-1,1.2)--(-1,0.3);
\node[anchor=east] at (-1,0.8){{\small $\langle \Phi_L  \rangle \sim ({\bf 1}, {\bf 14}, {\bf 1})$}};
\draw[->,thick] (1,1.2)--(1,0.3);
\node[anchor=west] at (1,0.8){{\small $\langle \Phi_R \rangle \sim  ({\bf 1}, {\bf 1}, {\bf 14})$}};
\end{tikzpicture}
\end{center}
\caption{{\small The symmetry breaking scheme in our model, as described in \S \ref{sec:SSB}. %At high scales $\Lambda_L$ and $\Lambda_R$ a pair of scalars condenses to break the electroweak-flavour-unified model down to an intermediate gauge theory, which features a deconstructed $SU(2)_L$ symmetry. At a lower scale $\Lambda_H$, the heavy components of the Higgs fields $H_1$ and $H_{15}$ are integrated out. At a slightly lower scale again, indicated by $\epsilon \Lambda_H$, the $\Gint$ theory is broken by the vevs of two more scalars down to the SM. The quantum numbers of all these scalars are recorded in Table~\ref{tab:Scalars}.  
}} \label{fig:breaking}
\end{figure}

The UV symmetry $G_{\mathrm{EWF}}$ is broken sequentially down to $G_{\mathrm{SM}}$, as summarized in fig.~\ref{fig:breaking}. At the high scale, the gauge-flavour-unified symmetries are first {\em deconstructed} to products of family specific symmetries,\footnote{
We remark that similar family-deconstructed gauge symmetries have been used in flavour model-building~\cite{Bordone:2017bld}, to explain the recent anomalies in semileptonic $B$ meson decays together with fermion masses. 
}
\begin{align}
&\langle S_L \rangle : Sp(6)_L &&\longrightarrow SU(2)_{L,1} \times SU(2)_{L,2} \times SU(2)_{L,3}\, , \\
&\langle S_R \rangle : SU(4)\times Sp(6)_R &&\longrightarrow SU(3) \times Sp(4)_{R,12} \times U(1)_R\, ,
\end{align}
where $U(1)_R$ acts as $B-L$ on the light right-handed fermions, and acts as hypercharge on everything else.
In the `right sector' the first two families remain unified. Also note that the scalar $S_R$ simultaneously breaks the quark-lepton unification.
The result of these high scale symmetry breakings is an intermediate gauge symmetry
\begin{equation} \label{eq:Gint}
G_{\mathrm{int}} = SU(3) \times \prod_{i=1}^3 SU(2)_{L,i} \times Sp(4)_{R,12} \times U(1)_R\, .
\end{equation}

Upon breaking to this intermediate symmetry, the fundamental Higgs fields $H_1$ and $H_{15}$ split into a set of flavour-specific components, with a component that couples to every pair of SM families (one left-handed, one right-handed). While we save a proper investigation of the scalar potential for future work, it is reasonable for the Higgs vev to fall into a small number of these family-aligned components, which we take to be the components $\mathcal{H}_a \sim [{\bf 1, (1, 1, 2), 1}]_{3}$ and $\overline{\mathcal{H}}_a \sim [{\bf 1, (1, 1, 2), 1}]_{-3}$, for each of $a=1,\, 15$.

We assume all other components of the fundamental Higgs fields are heavy, and are integrated out at a high scale $\Lambda_H \lesssim \Lambda_{L,R}$. The fact that the light Higgs components transform under one $SU(2)_{L,i}$ factor singles out one family of fermions that couples directly to the Higgs via the fundamental Yukawa interaction (\ref{eq:yuk}). This {\em defines} the third family (until this point, the theory is invariant under permuting the family labels of left-handed fermions). The other fermions are massless at the renormalisable level.

Lastly, the remaining two scalar fields $\Phi_L$ and $\Phi_R$ break the intermediate symmetry (\ref{eq:Gint}) down to the SM. These scalars start life in 2-index antisymmetric irreps of $Sp(6)_L$ and $Sp(6)_R$ respectively, which means that they decompose into sets of `link fields' under  (\ref{eq:Gint}). 
In the case of $\Phi_L$, which is a real scalar,
\begin{equation}
\Phi_L \sim {\bf 14}_L \mapsto {\bf 1}^{\oplus 2} \oplus ({\bf 2, 2, 1}) \oplus ({\bf 2, 1, 2}) \oplus ({\bf 1, 2, 2})\, 
\end{equation}
under (\ref{eq:Gint}),
where we only record the representations under $SU(2)_{L,i}$ factors on the RHS.
We thus get a link field charged under each pair of $SU(2)_{L,i}$ symmetries. Let $\phi_L^{ij}$ denote the link field transforming as a bidoublet of $SU(2)_{L,i}\times SU(2)_{L,j}$. If any two of these three link fields acquire non-zero vevs, then the deconstructed electroweak symmetry is broken to the diagonal,
\begin{equation}
\langle \Phi_L \rangle : SU(2)_{L,1} \times SU(2)_{L,2} \times SU(2)_{L,3} \longrightarrow SU(2)_L\, 
\end{equation}
which is the SM electroweak symmetry. 

We take the vevs to be in the $\phi_L^{12}$ and $\phi_L^{23}$ components (while we could have chosen otherwise, this is the choice that naturally reproduces the observed hierarchy in the three CKM angles),
with magnitudes of order
\begin{equation} \label{eq:phiL}
\langle \phi_L^{12} \rangle \sim \epsilon_L^{12} \Lambda_H\, , \qquad 
\langle \phi_L^{23} \rangle \sim \epsilon_L^{23} \Lambda_H\, .
\end{equation}
These vevs can be taken as definitions of the real parameters $\epsilon_L^{ij} <1$, which are ratios of scales.
We will see (\S \ref{sec:EFT}) that mild ratios between scales, $\epsilon_L^{12} \sim 1/5$ and $\epsilon_L^{23} \sim (1/5)^2$, are needed to explain the observed hierarchies in the SM Yukawas.

In the case of $\Phi_R$, which is a complex scalar,
\begin{equation}
\Phi_R \sim {\bf 14}_R \mapsto {\bf 1} \oplus {\bf 5}_0 \oplus {\bf 4}_{-3} \oplus {\bf 4}_3 \, 
\end{equation}
under (\ref{eq:Gint}),
where we only record the representations under $Sp(4)_R \times U(1)_R$ on the RHS. The ${\bf 5}$ is the antisymmetric 2-index irrep of $Sp(4)_R$ and serves as the link field $\phi_R^{12}$ for the first two families, while the ${\bf 4}_{\pm 3}$ components $\phi_R^{13,23}$ link the light families with the third. Suitably chosen vevs in all these components (see ref.~\cite{Davighi:2022fer} for the details) achieve the breaking
\begin{equation}
\langle \Phi_R \rangle :Sp(4)_R \times U(1)_R \longrightarrow U(1)_Y\, ,
\end{equation}
and we are left with the SM. As for the `left-sector', we introduce parameters $\epsilon_R^{12}$ and $\epsilon_R^{23}$, which set the vevs of $\phi_R^{12}$ and $\phi_R^{23}$ respectively in units of the heavy scale $\Lambda_H$.%\footnote{The formulae are not quite as simple as eq.~\ref{eq:phiL} in this case, because the fields $\phi_R^{12,23}$ are complex and allow independent components that couple to up-type and down-type components.}

\subsection{Effective field theory for the light Yukawas} \label{sec:EFT}

Recall that the light Higgs fields $\mathcal{H}_a$ and $\overline{\mathcal{H}}_a$ have renormalisable Yukawa couplings only to one family of fermions, which defines the third. Effective Yukawa couplings for all the light flavours are in fact generated in the steps described in \S \ref{sec:SSB}, with in-built hierarchies and enough parametric freedom to account for the fermion mass and quark mixing data. 

\begin{figure}[t]
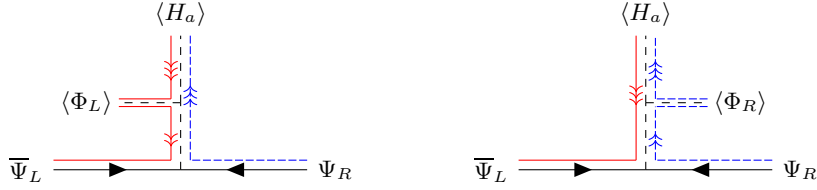

\begin{center}
\colordiagram{
    \fermions
    \HiggsII{L3}{R1}
    \phiLI{c1}
    \LII{L0}{L1}
    \LIII{L2}{L3}
    \RIII{R0}{R1}
} \qquad
\colordiagram{
    \fermions
    \HiggsII{L1}{R3}
    \phiRI{c1}
    \LIII{L0}{L1}
    \RII{R0}{R1}
    \RIII{R2}{R3}
}
\caption{{\small Feynman diagrams that contribute to the $Y_{23}$ (left) and $Y_{32}$ (right) elements of the Yukawa matrices. %, once the heavy Higgs component running along the internal lines are integrated out at $\Lambda_H = M_H$. 
The  \textcolor{red}{red} and \textcolor{blue}{blue} lines depict the flow of family indices embedded in \textcolor{red}{$Sp(6)_L$} and \textcolor{blue}{$Sp(6)_R$} respectively.
We see that $\Phi_L$ plays the role of a `link field', converting a third family left-index to a second family left-index.
}
\label{fig:dim5}}
\end{center}
\end{figure}

We here illustrate the mechanism with the simplest example, which is the generation of the effective $Y_{23}$ Yukawa couplings. (The full details of the effective field theory (EFT) matching are laborious, so we refer to ref.~\cite{Davighi:2022fer}.) Let us assume the presence of the following cubic scalar interaction in the UV,
\begin{equation} \label{eq:potential}
V(\phi) \supset \beta_L^1 \Lambda_H\, \mathrm{Tr}\left(\Omega^T H_1^\dagger \Omega \Phi_L \Omega H_1 \right)\, .
\end{equation}
Then, upon breaking $G_{\mathrm{EWF}} \to G_{\mathrm{int}}$ and integrating out the heavy Higgs components, the diagram on the left of fig.~\ref{fig:dim5}
generates dimension-5 operators
\begin{equation}
\mathcal{L}_{\mathrm{int}} \supset \frac{y_1 \beta_L^1}{2\Lambda_H}  \phi_L^{23} \left( \overline{Q}_2 \mathcal{H}_1 D_3 + \overline{Q}_2 \overline{\mathcal{H}}_1 U_3 + \overline{L}_2 \mathcal{H}_1 E_3 \right)
%\mathcal{L}_{\mathrm{int}} \supset \frac{y_1 \beta_L^1}{\Lambda_H} \mathrm{Tr}\left( \overline{\Psi}_L \Omega P_{12} \Phi_L \Omega P_3 H_1 \Omega^T \Psi_R \right)
\end{equation}
in the intermediate effective theory.
%EFT governed by symmetry (\ref{eq:Gint}), where the projection operators $P_3$ and $P_{12}$ project fields onto family-specific components. 
These dimension-5 operators then contribute to the Yukawa couplings $Y_{23}$ after $\Phi_L$ acquires its vev (there are also contributions from the $\mathcal{H}_{15}$ and $\overline{\mathcal{H}}_{15}$ Higgses), {\em i.e.} upon breaking $G_{\mathrm{int}} \to G_{\mathrm{SM}}$. 
The effective Yukawa couplings $Y_{32}$ are also generated at dimension-5, this time by operators involving $\Phi_R$ link fields (fig.~\ref{fig:dim5}, right), which require cubic interactions between $H_a$ and $\Phi_R$.

%The physical Higgs is the $P_3 H_1$ component, and the contraction with $P_{12} \Phi_L$ uniquely picks out the $\phi_L^{23}$ component of the vev. This couples to a second generation index in the fermion field $\overline{\Psi}_L$. 

% in fig.~\ref{fig:dim5}. We see that the $\phi_L^{23}$ link field converts a third family left-handed index to a second family left-handed index. The result of this diagram, after matching to the SM, is a set of Yukawa couplings of the form $\frac{\langle \phi_L^{23} \rangle}{\Lambda_H}\overline{\psi}_{L,2} \mathcal{H} \psi_{R,3}$, for each type of SM fermion. 

Continuing to higher dimensions in the $G_{\mathrm{int}}$-invariant EFT, the Yukawa couplings $Y_{13}$, $Y_{22}$, and $Y_{31}$ originate from dimension-6 operators involving {\em two} insertions of the symmetry breaking scalars $\Phi_{L,R}$. The $Y_{12}$ and $Y_{21}$ Yukawa couplings arise at dimension-7, and lastly the $Y_{11}$ couplings at dimension-8.  The result of all these contributions is the following hierarchical texture for each fermion mass matrix,
\begin{align} \label{eq:Mmatrix}
\frac{M^f}{v}
\sim \begin{pmatrix} 
\epsilon_L^{12}\epsilon_L^{23}\epsilon_R^{12}\epsilon_R^{23} &\epsilon_L^{12}\epsilon_L^{23}\epsilon_R^{23} &\epsilon_L^{12}\epsilon_L^{23}\\
\epsilon_L^{23}\epsilon_{R}^{12}\epsilon_{R}^{23} &\epsilon_L^{23}\epsilon_{R}^{23} &\epsilon_L^{23}\\
\epsilon_{R}^{12}\epsilon_{R}^{23} &\epsilon_{R}^{23} &1\\
\end{pmatrix}\, ,
\qquad
f = u, d, e\, .
\end{align}
Each entry in these matrices is accompanied by a coefficient which is some explicit function of the fundamental dimensionless couplings of the model, namely, the fundamental Yukawa couplings (\ref{eq:yuk}) plus the parameters in the scalar potential (only one of which we have written explicitly in eq.~\ref{eq:potential}).

The physical observables, {\em i.e.} the fermion masses and quark mixing angles, can then be extracted using matrix perturbation theory. They depend on the scale ratios $\epsilon_{L,R}^{ij}$ as
\begin{align}
y_{u,d,e} &\sim \epsilon_L^{12} \epsilon_R^{12} \epsilon_L^{23} \epsilon_R^{23}\, , \\
y_{c,s,\mu} &\sim \epsilon_L^{23} \epsilon_R^{23}\, , \\
y_{t,b,\tau} &\sim 1\, , \\
V_{us} &\sim \epsilon_L^{12}\, , \\
V_{cb} &\sim \epsilon_L^{23}\, , \\
V_{ub} &\sim \epsilon_L^{12}\epsilon_L^{23}\, .
\end{align}
The hierarchies in the CKM angles then favour scale suppressions of the order
\begin{equation}
\epsilon_L^{12} \sim \lambda \approx 0.23, \qquad \epsilon_L^{23} \approx \lambda^2,
\end{equation}
where $\lambda$ is the Cabibbo angle. The remaining $\epsilon_R^{ij}$ ratios can then be adjusted to reproduce the average mass hierarchies between the three generations. Since no pair of symmetry breaking scales are separated by more than an order of magnitude, it is expected that this ladder of scales is at least radiatively stable~\cite{Allwicher:2020esa}.
Finally, and crucially, it is shown in~\cite{Davighi:2022fer} that there is enough freedom in the particular EFT coefficients to actually fit all the data (beyond just reproducing the observed hierarchies), despite some interrelations between elements of the CKM matrix. %The nine fermion masses can be fit as independent $\mathbb{C}$-numbers, as can the CKM elements $V_{us}$, $V_{cb}$, and $V_{ub}$, with all other CKM elements being then fixed 

\subsection{Electroweak Flavour Unification: Outlook}

To summarize, the electroweak flavour unification model offers a new explanation for the origin of three generations of fermions and the hierarchical structure of fermion masses and quark mixing angles, in terms of a flavour-enriched version of the quark-lepton unifying symmetry of Pati and Salam. 
We emphasize that no extra ingredients, such as vector-like fermions, are required to generate the Yukawa structure, beyond the scalar fields needed to break $G_{\text{EWF}}$ down to the SM.

Beyond reproducing these known facts of Nature, we have not discussed the phenomenological consequences of such a framework -- which, if the unification  scales are not too far off, would be multitudinous. Indeed, the $SU(4) \times Sp(6)_L \times Sp(6)_R$ gauge model does not give rise to proton decay, and so the mass scale of the heavy gauge bosons can in principle be low. The lightest gauge bosons in the model will be those arising from the final breaking steps, $G_{\mathrm{int}} \to G_{\mathrm{SM}}$, which transform as a selection of {\em flavoured} $W^\prime$ and $Z^\prime$ bosons, some coupled to left-handed fermions and others to right-handed fermions. We save an explanation of the low-energy phenomenology signatures of the model, which are complicated, for future work. For now, we remark only that the extra states will provide extra sources of quark flavour violation and lepton universality violation, for which there is intriguing evidence already in the decays of $B$ mesons.

\section{Postscript: colour flavour unification} \label{sec:colour}

We conclude this proceeding with some comments about the other route to unifying gauge and flavour symmetries, namely `colour flavour unification' based on the (admittedly rather large) gauge group $G_{\text{CF}} = SU(12) \times SU(2) \times SU(2)$~\cite{Davighi:2022qgb}. The key observation we wish to make is that there is a sequence of subgroup embeddings,
\begin{equation}
G_{\text{SM}} \times \mathbb{Z}_9 \hookrightarrow G_{\text{SM}} \times U(1)_{\text{LFUV}} \hookrightarrow G_{\text{CF}}\, .
\end{equation}
Here the discrete symmetry $\mathbb{Z}_9$ acts on all quarks with a charge of 1 mod 9, and on leptons with a charge of 0 mod 9. A family of possible $U(1)_{\text{LFUV}}$ symmetries that fit in this sequence are generated by the following combinations of baryon number and lepton family numbers
\begin{equation}
%X = 3(3a+r)B + 9(b-a)(L_e +L_\tau) -9(a+2b+r)L_\mu\, , 
X = 3B + 18 r(L_e +L_\tau) -9(4r+1)L_\mu\, , 
\end{equation}
where $r\in \mathbb{Z}$. These $U(1)$ symmetries violate lepton flavour universality, hence the label `LFUV'.\footnote{In~\cite{Davighi:2022qgb}, building on refs.~\cite{Davighi:2020qqa,Greljo:2021xmg,Greljo:2021npi}, these $U(1)_{\text{LFUV}}$ symmetries were used to justify why scalar leptoquarks at the TeV scale might have `muophilic' couplings, to explain the evidence for new physics in muons (notably, the $B$ decay anomalies and the anomaly in $(g-2)_\mu$) without also causing charged lepton flavour violation.}

Interestingly, the symmetry breaking $\pi:U(1)_{\text{LFUV}} \to \mathbb{Z}_9$ can be triggered by the vevs of a pair of scalars $\phi_1$ and $\phi_2$ with precisely the right $U(1)_{\text{LFUV}}$ charges, namely 
$q_1=36r$ and $q_2=9(2r+1)$,
%$q_1=18(a-b)$ and 
%$q_2=9(2a+b+r)$, 
needed to sufficiently populate (up to a single zero entry) the $3 \times 3$ neutrino Majorana matrix via renormalisable couplings $\sim \phi_a \overline{\nu}^c_i \nu_j$. %, in order to account for all data pertinent to neutrino masses and mixings. 
Thus, if the breaking $\pi$ occurs at the traditional see-saw scale, of order $10^{14}$ GeV, then realistic neutrino masses and mixings are generated, and one is left with the gauged $\mathbb{Z}_9$ symmetry that acts only on quarks (and thence on baryons, thence nuclei). This remnant $\mathbb{Z}_9$ gauge symmetry persists all the way to the deep IR. 

While the existence of this discrete gauge symmetry would likely evade positive confirmation by any conventional experiment, it can certainly be falsified because it leads to the strict selection rule,
\begin{equation}
\Delta B = 0 \mathrm{~mod~} 3\, ,
\end{equation}
which would be {\em exact} to all orders in the SM EFT. Protons are therefore exactly stable in a model with this $\mathbb{Z}_9$ symmetry, regardless of whatever new physics might be lurking in the UV (be it colour flavour unification, or anything else). Nor do neutrons oscillate, for example, since $\Delta B =2$ processes are forbidden. High scale leptogenesis {\em is} possible, on the other hand, because the $\Delta B = 3$ electroweak sphaleron is allowed.

The fact that such a `fine' $\mathbb{Z}_9$ discrete gauge symmetry, under which the leptons are neutral, emerges as a remnant in the IR is a direct consequence of gauge flavour unification in the UV. If there were only one generation, then adapting the preceding argument (starting this time from the Pati--Salam symmetry) gives us at most a mod 3 quark symmetry, from breaking $B-L$. This gives no protection of baryon number, and indeed forbids nothing that is not already forbidden by QCD. If there were two generations, then colour flavour unification can spit out a mod 6 gauge symmetry capable of stabilizing the proton -- but it would not prevent neutrons from oscillating. It is only with three generations (at least) that both $\Delta B = 1$ and $\Delta B = 2$ processes are elegantly forbidden in this framework.
We arrive at the notion that the existence of multiple generations might be indirectly responsible for the stability of the proton.

%ROUGH...
%If there were only one family, then PS symmetry in the UV, and the most we can get out is a $\mathbb{Z}_3$ discrete symmetry (from breaking $B-L$) that gives no baryon number protection. If there were two families and colour flavour unification in the UV, the $\mathbb{Z}_6$ would stabilize protons but not prevent neutrons from oscillating. It is only with 3 families (at least) that both $\Delta B = 1$ and $\Delta B = 2$ processes are elegantly forbidden in this framework.
%Connect this with LFUV and gauge-flavour unification. The fact that the existence of many flavours might be intimately connected to the stability of protons is an intriguing new connection that might be worth further exploration.

\section*{Acknowledgments}
I am very grateful to Joseph Tooby-Smith for collaboration on the work~\cite{Davighi:2022fer} on which this talk is predominantly based, and to Admir Greljo and Anders Eller Thomsen for collaboration on~\cite{Davighi:2022qgb}. I thank the organizers of La Thuile 2022 for the invitation to present this work, and for organizing a wonderful conference.
I am supported by the SNF under contract 200020-204428, and by the European Research Council (ERC) under the European Union’s Horizon 2020 research and innovation programme, grant agreement 833280 (FLAY).

\end{document}